\documentclass[final]{ptptex}
\usepackage{graphicx}
\usepackage{amsmath}
\usepackage{amssymb}
\usepackage[normalem]{ulem}  
\usepackage[dvips]{color} 

\renewcommand\sout{\bgroup \color{red} \ULdepth=-.5ex \ULset}

\newcommand{\Slash}[1]{{\ooalign{\hfil/\hfil\crcr$#1$}}}
\newcommand{\skipthis}[1]{}

\begin{document}
 \title{Functional Renormalization Group Study of Phonon Mode Effects on Chiral Critical Point}

 \author{Kazuhiko
 \textsc{Kamikado}$^{1}$\footnote{kamikado@yukawa.kyoto-u.ac.jp},
 Teiji \textsc{Kunihiro}$^{2}$,
 Kenji \textsc{Morita}$^{1}$ and Akira \textsc{Ohnishi}$^{1}$}

 \inst{$^1$Yukawa Institute for Theoretical Physics, Kyoto University,
 606-8502, Kyoto,  Japan,\\
 $^2$Department of Physics, Kyoto University, 606-8502, Kyoto, Japan} 


 \maketitle

  \begin{abstract}%
   We apply the functional renormalization-group (FRG) equation to
   analyze the nature of the QCD critical point beyond the mean-field
   approximation by taking into consideration the fact that the soft
   mode associated with the QCD critical point is a linear combination
   of fluctuations of the chiral condensate and the quark-number
   density, rather than the pure chiral fluctuations.  We first
   construct an extended quark-meson model in which a new field
   $\varphi$ corresponding to quark-number density is introduced to the
   conventional one composed of the chiral fields $(\sigma, \pi)$ and
   the quarks.  The fluctuations of the quark-number density as well as
   the chiral condensate are taken into account by solving the FRG
   equation which contains $\sigma$ and $\varphi$ as coupled dynamical
   variables.  It is found that the mixing of the two dynamical
   variables causes a kind of level repulsion between the curvature
   masses, which in turn leads to an expansion of the critical region of
   the QCD critical point, depending on the coupling constants in the
   model yet to be determined from microscopic theories or hopefully by
   experiments.
  \end{abstract} 
  \maketitle
 \section{Introduction}
 The phase diagram of Quantum Chromodynamics (QCD) has been one of the
 main topics of theoretical high energy nuclear physics. In fact QCD
 phase diagram is expected to have a rich structure in a temperature
 ($T$) and quark-chemical potential ($\mu$) plane
 \cite{BraunMunzinger:2009zz,Friman:2011zz,Fukushima:2010bq}.  At low
 temperature and low chemical potential, the matter is considered to be
 strongly correlated and a naive perturbation does not work.  Therefore
 non-perturbative methods are necessary to analyze the phase structure.

 Lattice Monte-Carlo simulation is a first-principle and powerful
 approach for non-perturbative properties of QCD. At finite
 quark-chemical potential, however, we encounter the notorious sign
 problem which makes the statistical importance sampling method
 difficult. There are various proposals to overcome the sign problem
 \cite{deForcrand:2010ys}, but the cold and dense region $\mu \gg T$ is
 still inaccessible by lattice simulations.

 On the other hand, approaches based on effective models embodying
 proper symmetries with low-energy effective or collective degrees of
 freedom are legitimate and useful methods to investigate the phase
 diagram where we need to consider only low-lying soft modes with the
 energy scale of temperature $\sim 200$ MeV.  Many investigations of the
 QCD phase diagram have been done on the basis of effective models,
 including (Polyalov loop extended-)Nambu-Jona-Lasinio ((P)NJL) model
 \cite{Asakawa:1989bq,Hatsuda:1994pi,Buballa:2003qv,Fukushima:2003fw,Ratti:2005jh}
 or (Polyalov loop extended-) Quark-meson ((P)QM) model
 \cite{Braun:2003ii,Jungnickel:1995fp,Schaefer:2007pw}.  Whereas there
 is a strong model dependence of the position of the critical point
 \cite{Stephanov:2004wx}, many chiral effective model calculations
 suggest the existence of a QCD critical point at finite temperature and
 quark-chemical potential where the crossover transition in the
 lower-$\mu$ region becomes a second order phase transition followed by
 a first-order in the higher-$\mu$ region. 

 In the vicinity of the QCD critical point, there is a critical
 region in which susceptibilities \cite{Hatta:2002sj} or their higher
 derivatives \cite{Stephanov:2008qz} show critical behavior. In
 experimental searches of the QCD critical point, it is helpful to
 examine a shape and size of the critical region as discussed in
 Refs.~\cite{Schaefer:2006ds,Costa:2008gr,Fu:2010ay,Schaefer:2011ex}.
   
 In general, there should exist at least one softening mode inherent in
 the second-order nature of the transition around a second order phase
 boundary. Analyses of the dynamical soft mode have been done on the
 basis of the quantum field theory \cite{Fujii:2003bz,Fujii:2004jt} and
 the Langevin equation \cite{Son:2004iv}. Both analyses suggest that the
 soft mode around the critical point should be the quark-number density
 fluctuation or phonon mode (particle-hole excitation), in addition to
 the chiral order parameter.  Since the quark number is a
 conserved quantity, its fluctuating mode ( phonon or sound mode) is
 necessarily a soft mode in the sense that its excitation energy
 $\omega$ vanishes in the small wave-number ($k$) limit as $\omega =c_s
 k$ where $c_s$ is the sound velocity. Then an effective potential as a
 function of the baryon-number density in addition to the chiral order
 parameter should be applied to investigate the critical phenomena
 accompanying the critical point \cite{Fujii:2004jt,Son:2004iv}. This
 situation is identical to the first-order liquid-gas phase transitions
 of water and the nuclear matter. As known in the $\sigma$-$\omega$
 model \cite{Walecka:1974qa,Serot:1984ey}, the chiral order parameter
 couples with the baryon-number density mode at finite chemical
 potential. Consequently the linear combination of these two modes would
 be realized as the soft mode around the critical point. Although the
 inclusion of the quark-number density mode does not change the
 universality class, i.e., the critical point has the same universality
 class with the Ising model in three dimensions, it can change
 quantitative features, such as the size or shape of the critical
 region. It is worth emphasizing that they are of phenomenological
 importance, since an extension of the critical region toward higher-$T$
 and lower-$\mu$ direction, which is found to be the case in the present
 work, implies that we may have more chance to probe the critical
 behavior by heavy-ion collision experiments.
 
 Recently alternative possibilities of the phase structure are
 suggested, for example, inhomogeneous chiral condensation phases with
 the Lifshitz point \cite{Nakano:2004cd,Nickel:2009wj,Carignano:2010ac}
 or multiple chiral critical points
 \cite{Kitazawa:2002bc,Yamamoto:2007ah,Zhang:2008wx,Zhang:2010qe,Zhang:2011xi}.
 In the present work, however, we focus on the conventional phase
 structure with a single first-order phase boundary at low temperatures
 and a single critical point.  Since the coupling between the
 quark-number density and the chiral condensate is inevitable at finite
 chemical potential, the analysis in the present work is also
 appropriate for the other candidates of the phase structure.
 
 Near the second-order phase boundary, fluctuations of the soft mode are
 enhanced and the mean-field approximation breaks down.  We need to
 include the soft mode and its fluctuation effects in order to correctly
 describe the QCD critical point and its critical region. The functional
 renormalization group (FRG) method
 \cite{Polchinski:1983gv,Wegner:1972ih,Wetterich:1992yh} is one of the
 machineries to evaluate an effective action beyond the mean-field
 approximation, see \cite{Berges:2000ew,Pawlowski:2005xe} for a general
 introduction.  Although there are many analyses on the QCD phase
 diagram using FRG
 \cite{Berges:1998sd,Schaefer:2006ds,Nakano:2010PLB,Skokov:2011PRC,Herbst:2011PLB,Morita:2011PRD,Braun:2011PRL}
 where effective models are mostly used, there have been only few works
 on the QCD critical point which include the quark-number density
 fluctuation as well as chiral modes beyond the mean-field
 approximation.

 In the present work, we start from an effective model including the
 quark-number density fluctuation as well as the sigma and pion modes.
 We apply the FRG method to take into account the fluctuations of the
 soft mode.  A key quantity in our analysis is a strength of the
 coupling between the sigma and quark-number density.  We thereby
 examine not only the phase structure around the critical point but also
 the sensitivity of the critical region to the sigma-density coupling
 strength.

 This paper is organized as follows. In Sec. \ref{sec:Formulation}, an
 evolution equation for the scale dependent effective potential is
 derived.  In Sec. \ref{sec:effective-model-near}, we introduce an
 effective model which should describe the dynamics near the QCD
 critical point.  The new field $\varphi$ is introduced in order to
 describe the quark-number density. The FRG method and a suitable
 approximation for our model is discussed in
 Sec. \ref{sec:funct-renorm-group}. Results of numerical calculations
 are given in Sec. \ref{sec:numerical-results}. Section
 \ref{sec:summary} is devoted to the summary and outlook.
   
 \section{Formulation} \label{sec:Formulation}
  \subsection{Effective model near QCD critical point}
  \label{sec:effective-model-near} In this section we introduce an
  effective model for describing the two-flavor QCD near the critical
  point. Our starting point is the quark-meson (QM) model
  \cite{Jungnickel:1995fp} which is one of the chiral effective models
  of the low-energy QCD.  Since the QM model explicitly contains the
  meson degrees of freedom in the Lagrangian, we can easily incorporate
  the fluctuations of the mesonic modes in the framework of the FRG.
  The Lagrangian of the QM model for 2-flavor and 3-color degrees of
  freedom reads
  \begin{equation}
   \begin{split}
    {L}_{\rm QM} &= \bar{\psi} [ i\Slash{\partial} - g_s (\sigma + i
    \gamma_5
    \vec{\tau} \cdot \vec{\pi}) ] \psi 
    +\frac{1}{2}(\partial_{\mu} \sigma)^2+\frac{1}{2}(\partial_{\mu}
    \vec{\pi})^2  -U(\sigma^2+\vec{\pi}^2) + c \sigma,
    \label{eq:Lagransian_QM}
   \end{split} 
  \end{equation}
  with the mesonic potential $U(\sigma^2 + \vec{\pi}^2) =
  \frac{m^2}{2}(\sigma^2 + \vec{\pi}^2) + \frac{\lambda}{4!}(\sigma^2 +
  \vec{\pi}^2)^2$.  The explicit symmetry breaking term ($c \sigma$)
  corresponds to a finite current quark mass.  Except the explicit
  symmetry-breaking term, the Lagrangian is invariant under the chiral
  $SU(2)_L \times SU(2)_R$ transformation.

  The QM model has been extensively used as the starting model for the
  FRG analysis of the global structure of the QCD phase diagram
  \cite{Braun:2003ii,Jungnickel:1995fp,Schaefer:2007pw}, i.e., a chiral
  symmetry breaking at low temperature and low chemical potential and
  its restoration at high temperature and/or high chemical
  potential. Many previous works using the QM model rely on the analysis
  of the effective potential as a function of the chiral order parameter
  ($\sigma$) : Notice that the $\sigma$ field in
  Eq. (\ref{eq:Lagransian_QM}) couples to the quark scalar density
  $\bar{\psi}\psi$, the expectation value of which is the primary chiral
  order parameter. In those analyses, the finite quark density effects
  are taken into account through the chemical potential dependence of
  the effective potential. As a result, the critical behaviors
  associated with the QCD critical point were caused only by the
  fluctuation of the chiral order parameter or scalar density. The soft
  mode associate with the QCD critical point is, however, a linear
  combination of the chiral order parameter and the quark-number density
  \cite{Fujii:2003bz,Fujii:2004jt,Son:2004iv}, which has not been
  properly taken into account in the previous works based on the FRG
  analysis despite of its importance.  Thus an FRG analysis using the
  effective potential as a function of not only the chiral order
  parameter but also the quark-number density as an independent
  additional degree of freedom is of imperative importance to understand
  the phase structure near the critical point precisely.
  
  In this work in order to effectively incorporate the
  fluctuation of the quark-number density, we extend the QM model. We
  introduce a new bosonic degree of freedom $\varphi$ following the
  analyses in Refs.~\cite{Fujii:2004jt} and \cite{Son:2004iv}: the new
  field $\varphi$ corresponds to the quark-number density with an
  appropriate normalization and naturally couples to $\bar{\psi}\gamma_0
  \psi$. Thus our effective model takes the following form:
  \begin{equation}
   \begin{split}
    L &= \bar{\psi} [ i\Slash{\partial} - g_s (\sigma + i \gamma_5
    \vec{\tau} \cdot \vec{\pi}) + g_d \varphi \gamma_{0} ] \psi 
    +\frac{1}{2}(\partial_{\mu} \sigma)^2+\frac{1}{2}(\partial_{\mu}
    \vec{\pi})^2 +\frac{1}{2}(\partial_{\mu}
    {\varphi})^2  \\
    & -U(\sigma^2+\vec{\pi}^2)-\frac{ m_{\varphi}^{2}}{2}\varphi^2 + c
    \sigma.
    \label{eq:Lagransian_dQM}
   \end{split} 
  \end{equation}

  Whereas the effective model (\ref{eq:Lagransian_dQM}) looks similar to
  the NJL model with a vector interaction
  \cite{Kunihiro:1991qu,Kitazawa:2002bc,Sasaki:2006ws,Sakai:2008ga,Fukushima:2008wg,Fukushima:2008is},
  our model has only the mode corresponding to the zeroth component of
  the vector fields. One might suspect that the other components of the
  vector fields should be included in order to respect the manifest
  Lorentz invariance.  It should be, however, noted that we apply the
  model to finite chemical potential where baryonic matter is formed and
  the Lorentz symmetry is explicitly broken.  Moreover the other
  components of the vector fields are irrelevant to the critical
  phenomena near the QCD critical point because we do not have finite
  condensations of the spatial components and then the spatial
  components is decoupled from the chiral modes.  In this sense, our
  model embodies enough ingredients for the investigation of the
  critical phenomena associate with the QCD critical point.

  There also exists a difference in the nature between the $\varphi$ and
  the zeroth component of the vector fields.  In our set up, the field
  $\varphi$ which is supposed to describe a particle-hole excitation and
  causes an attractive interaction between two quarks
  \cite{Fetter:1971ev} while the zeroth component of the vector field is
  usually introduced as a source of repulsive interaction. We will come
  to this point later.

  At finite chemical potential, both the chiral condensate and
  quark-number density take finite expectation values, $\langle \sigma
  \rangle \neq 0$ and $\langle \varphi \rangle \neq 0$.  We calculate
  the effective potential with the fluctuations of the chiral modes and
  $\varphi$ by solving the FRG equation and determine the expectation
  values of these modes from the minimum of the effective potential.
       
  \subsection{Functional renormalization group}
  \label{sec:funct-renorm-group} In order to evaluate the effective
  potential with the fluctuations of the quark-number density as well as
  the chiral modes, we adopt the FRG equation. We make an approximation
  suited to solve the QM model and derive an evolution equation for a
  scale dependent effective potential.

  In the FRG approach, we introduce a scale ($k$) dependent effective
  action $\Gamma_k$ which should satisfy the following properties at
  ultraviolet ($k=\Lambda$) and infrared ($k=0$) scale:
  \begin{equation}
   \begin{split}
    \Gamma_\Lambda &= S \\ \Gamma_ 0 &= \Gamma \;,
    \label{eq:condition_gamma_k}
   \end{split}
  \end{equation}
   where $S$ and $\Gamma$ are the classical action and quantum effective
   action, respectively.
   
   We can construct the scale dependent effective action by inserting a
   mass-like term into the Lagrangian which controls the quantum and
   thermal contribution to the effective action. For the bosonic modes,
   we insert the bosonic mass-like term $R_{kB}(p)\phi(p)\phi(-p)$ into
   the Lagrangian where $R_{kB}(p)$ is an arbitrary function satisfying
   the following condition:

   \begin{equation}
    \begin{split}
     R_{kB} (p) \sim k^2 \;\; {\rm for}\;\; p^2 \ll k^2 ,\\
     R_{kB} (p) \sim 0 \;\;{\rm for}\;\;   p^2 \gg k^2 .\\
     \label{eq:condition_R_k}
    \end{split}
   \end{equation}

   At the infrared scale ($k=0$), $R_k$ is zero for any momentum, and
   $\Gamma_k$ becomes the quantum effective action.  At the ultraviolet
   scale ($k=\Lambda$), all modes acquire large masses, $m^2 \to m^2 +
   \alpha \Lambda^2$ where $\alpha$ is a proportional coefficient in
   $R_k(p)$, $R_\Lambda(p) \simeq \alpha \Lambda^2$. As a result, $R_k$
   prevents the propagation of all the fluctuation, and $\Gamma_\Lambda$
   becomes the classical action.  Thus one finds that the conditions for
   the $\Gamma_k$ are satisfied.  At an intermediate scale, $R_k(p)$
   suppresses propagation of the modes whose momentum are smaller than
   $k$.  Thus we can interpret the $\Gamma_k$ as an effective action
   which contains the fluctuations whose momentum scales are larger than
   $k$.  The above procedure can be extended to the fermionic modes by
   introducing the mass-like term $\bar{\psi}(p) R_{kF}(p) \psi(p)$
   where $R_{kF}$ has Dirac indices. (See, e.g.,
   Ref.~\cite{Berges:2000ew}).

   In the present work, we choose the so-called optimized cutoff
   function \cite{Litim:2001up} whose form is
   \begin{equation} 
    \begin{split}
     R_{k \rm B}(p)& = (k^2 - \vec{p}\;^2) \theta (k^2 - \vec{p}\;^2)
     \;\;\;{\rm for \;Bosons}\\
     R_{k \rm F}(p)& =
     \Slash{\vec{p}}\left(\sqrt{\frac{k^2}{\vec{p}\;^2}}-1 \right)
     \theta (k^2 - \vec{p}\;^2) \;\;\;{\rm for \;Fermions}\;.
     \label{eq:cutoff_function}
    \end{split} 
   \end{equation}

   The evolution of the $\Gamma_k$ with scale $k$ is described by the
   FRG equation \cite{Wetterich:1992yh}:
   \begin{equation}
    \begin{split}
     k\frac{\partial \Gamma_k }{\partial k}& = -{\rm Tr} \left[
     \frac{\partial_k R_{k \rm F}}{ R_{k \rm F} + \Gamma_k^{(2,0)}}\right]_{\rm F} 
     +\frac{1}{2}{\rm Tr} \left[
     \frac{\partial_k R_{k \rm B}}{ R_{k \rm B} +
     \Gamma_k^{(0,2)}}\right]_{\rm B} \;,
     \label{eq:exact_flow_equation}
    \end{split}
   \end{equation}

   where the trace runs over Dirac, color and flavor indices as well as
   momentum.  The superscript ${(a,b)}$ in $\Gamma_k^{(a,b)}$ denotes
   $a(b)$ times fermionic (bosonic) functional derivative of the
   $\Gamma_k$.  The first and second terms of r.h.s of
   (\ref{eq:exact_flow_equation}) are the fermionic and bosonic
   contributions respectively.

   To solve the FRG equation of the scale dependent effective action
   (\ref{eq:exact_flow_equation}), we need scale ($k$) dependencies of
   the 2-point functions $\Gamma_k^{(2,0)}$ and $\Gamma^{(0,2)}_k$.  In
   general, the FRG equation for the $n$-point function depends on up to
   the $(n+2)$-point function and the set of the FRG equations for
   $n$-point functions forms an infinite tower of the equations.  In
   practice, we introduce some approximations in order to close the
   tower of equations up to finite $n$-point functions.  In the present
   work, we apply the so-called local potential approximation (LPA) to
   the effective action.  We assume the form of the scale dependent
   effective action as
   \begin{equation}
    \begin{split}
     \Gamma_{k}^{LPA} &= \bar{\psi} [ i\Slash{\partial} - g_s (\sigma
     + i \gamma_5
     \vec{\tau} \cdot \vec{\pi}) - g_d \varphi \gamma_{0} + \mu \gamma_0 ] \psi 
     +\frac{1}{2}(\partial_{\mu} \sigma)^2+\frac{1}{2}(\partial_{\mu}
     \vec{\pi})^2 +\frac{1}{2}(\partial_{\mu}
     {\varphi})^2  \\
     & -U_k(\sigma^2+\pi_0^2,\varphi) - c \sigma \;,
     \label{eq:Local_potential_approximation}
    \end{split} 
   \end{equation}
   where $U_k$ is a scale dependent effective potential such that it
   agrees with a usual effective potential at $k=0$. Except for the
   small explicit breaking term $c \sigma$, which can be treated as a
   perturbation for the symmetric effective action, the
   Lagrangian~(\ref{eq:Lagransian_dQM}) respects the chiral symmetry.
   To a good approximation, we can regard $U_k$ as a function of
   $\sigma^2 + \vec{\pi}^2$ and $\varphi$.

   Here we neglect the scale dependence of the wave function
   renormalization for simplicity.  It would be important for the
   discussion of the possibility of the inhomogeneous condensation phase
   but it is out of the scope of this work.
   
   We also neglect the evolutions of the Yukawa coupling $g_s$ and the
   density coupling $g_d$. The inclusion of the scale dependencies of
   these couplings might change the global structure of the phase
   diagram such as the position of the QCD critical point or the slope
   of the first-order phase boundary of the chiral phase transition.
   Near the QCD critical point, however, the contributions from the
   light bosons are enhanced and the scale dependence of $g_s$ and $g_d$
   would not change the local phase structure, i.e., the shape or size
   of the critical region. Thus the scale dependence could be neglected
   when the critical region is only concerned.

   Now the 2-point functions in the r.h.s of
   (\ref{eq:exact_flow_equation}) is given by the functional derivatives
   of the LPA effective action:
   \begin{equation}
    \begin{split}
     \Gamma^{(2,0)}_k &\rightarrow \frac{\partial^2
     \Gamma_{k}^{LPA}}{\partial \bar{\psi} \partial \psi} \\
     \Gamma^{(0,2)}_{k\;a,b} &\rightarrow \frac{\partial^2
     \Gamma_{k}^{LPA}}{\partial \phi_a \partial \phi_b} \;.
     \label{eq:LPA_2ptfunction}
    \end{split}
   \end{equation}
   Using LPA, we can obtain an evolution equation for the scale
   dependent effective potential. Substituting the LPA ansatz
   (\ref{eq:Local_potential_approximation}),~(\ref{eq:LPA_2ptfunction})
   and the optimized cutoff function (\ref{eq:cutoff_function}) into
   (\ref{eq:exact_flow_equation}), we obtain the flow equation for the
   effective potential:
   \begin{equation}
    \begin{split}
     \frac{\partial U_k}{\partial k} &= \frac{k^4}{12 \pi^2}\Bigg[-2
     N_f
     N_c \left[\frac{1}{E_q} \tanh \left( \frac{E_q + (\mu +
     g_d \varphi)}{2T} \right) +  \frac{1}{E_q} \tanh \left( \frac{E_q - (\mu +
     g_d \varphi)}{2T} \right) \right]  \\
     &\;\;\;\;\;\;\;\;\;\;\;\;\;\;\;\;+ \sum_{i=\pm} \frac{1}{E_i} \coth\left(\frac{E_i}{2 T} \right) +
     \frac{3}{E_{\pi}} \coth \left(\frac{E_{\pi} }{2T}\right)
     \Bigg], \\
     \label{eq:LPA_flow_equation}
    \end{split}
   \end{equation}
   where the scale dependent particle energies $E_{\alpha}$ ($\alpha =
   q$, $\pi$ and $\pm$) are defined as
   \begin{equation}
    \begin{split}
     E_{\alpha}^2 &= k^2 + M_{ \alpha}^2, \\
     M_q^2 & =  g^2\sigma^2 ,\; M_{\pi}^2 =  \frac{U'}{\sigma}, \\
     M_{i = \pm}^2 &= \frac{ U'' + \ddot{U} \pm  \sqrt{(U'' - \ddot{U})^2 + 4 \dot{U}^{'2} }}{2}.
     \label{eq:masses}
    \end{split}
   \end{equation}
   Here we have used the following notation
   \begin{equation}
    U' = \frac{\partial U}{\partial \sigma} ,\;\;\dot{U}
     =\frac{\partial U}{\partial \varphi} \;.
   \end{equation}
   $M_{\pi}^2$ and $M_i^2$ are understood as the squared curvature
   masses of the $\pi$ and the linear combination of $\sigma$ and
   $\varphi$.

   Since the scale dependent effective action at the UV scale agrees
   with the classical action (\ref{eq:condition_gamma_k}), the initial
   condition of the flow equation (\ref{eq:LPA_flow_equation}) reads
   \begin{equation}
    \begin{split}
     U_{\Lambda}(\sigma^2+\vec{\pi}^2, \varphi) = a (\sigma^2 +
     \vec{\pi}^2) + b
     (\sigma^2 + \vec{\pi}^2)^2 + \frac{M^2_{\varphi}}{2} \varphi^2 \;.
    \end{split}
   \end{equation}
   By solving the flow equation from $k=\Lambda$ to $k = 0$, we obtain
   the effective potential with the fluctuation effects incorporated.
   The expectation values of the $\sigma$ and $\varphi$ are determined
   by minimizing the effective potential at $k=0$:
   \begin{equation}
    \begin{split} 
     \frac{\partial U}{\partial \sigma} \Bigg|_{\sigma_0,\varphi_0} - c =
     \frac{\partial U}{\partial \varphi}\Bigg|_{\sigma_0,\varphi_0}
     = 0\;.
     \label{eq:gap-equation}
    \end{split}
   \end{equation}
   We obtain the effective potential at various temperatures and
   chemical potentials, and the phase structure is determined by the
   chiral order parameter behavior.

   \section{Numerical results}
   \label{sec:numerical-results} In this section, we show numerical
   results. One of our concerns is the sensitivity of the phase
   structure to the strength of the mixing between the chiral condensate
   and the quark-number density, i.e., the magnitude of the density
   coupling $g_d$.

   First we calculate the phase diagrams with $g_d$ being varied. We
   show the position of the QCD critical point slightly moves to a
   higher temperature direction with increasing $g_d$.

   Next we describe the critical region which is associated with the QCD
   critical point. Finally we calculate the curvatures of the effective
   potential.  We show that a kind of level repulsion between the
   curvature masses occurs around the QCD critical point at finite
   $g_d$. This level repulsion causes an expansion of the critical
   region.

   \subsection{Calculation method and parameter fixing}
   \label{sec:calc-meth-param} We solve the flow equation
   (\ref{eq:LPA_flow_equation}) numerically on a grid in the field
   space. Since both $\sigma$ and $\varphi$ can take non-vanishing
   expectation values, we have to prepare the $U_k$ values on the
   two-dimensional grid of discrete field values in the $\sigma$ and
   $\varphi$ space.  Derivatives of the effective potential used in the
   numerical calculation are given by finite differences.  This reduces
   the flow equation to a coupled ordinary differential equations which
   can be integrated out with a standard method. We solve it by using
   the fourth-order Runge-Kutta method.

   There are few works using a 2-dimensional grid method for dynamical
   fields and, only very recently, it has been developed by several
   researchers including one of the present authors on the studies of
   QCD phase diagram with isospin chemical potential
   \cite{Svanes:2010we,Kamikado:2012bt} as well as the phase diagram of
   2-color QCD \cite{Strodthoff:2011tz}.  In comparison with the Taylor
   method in which the effective potential is locally evaluated around a
   scale dependent minimum, the grid method captures the global
   structure of the effective potential in the field space and is well
   suited to investigate the critical region with which the first order
   phase boundary is connected.
   \begin{table}[t]
    \begin{center}
     \begin{tabular}{|c|c|c|c|c||c|}
      \hline
      $a/{\Lambda^2}$ & $b$ & ${c}/{\Lambda^3}$&$M_{\varphi}^2 /\Lambda^2$ &$\Lambda$ MeV & $g_s$\\
      \hline
      0.3224&0.25 &0.0045 &0.1& 900 &3.2\\
      \hline
     \end{tabular}
    \end{center}
    \caption{Table of the initial parameters for the numerical
    calculations} \label{tab:parameter_frg}
   \end{table}
   In the present calculation, we choose the ultraviolet scale as
   $\Lambda= 900$MeV.  Our effective model has six parameters
   $a,b,c,g_s,M_{\varphi}$ and $g_d$.  The first four parameters have
   been determined to reproduce the vacuum physical values such as the
   pion mass, sigma mass, pion decay constant and constituent quark mass
   \cite{Kamikado:2012bt}.
   
   We can not, however, determine the remaining two parameters
   $M_{\varphi}$ and $g_d$ because at present any experimental data for
   the collective excitation corresponding to the particle-hole mode
   (phonon mode) in quark matter is not available.  In principle we
   could determine these parameters from microscopic theories such as
   QCD itself or the NJL model. However the main purpose of this work is
   to examine the sensitivity of the critical region to the mixing
   between the $\sigma$ and $\varphi$ then we treat $M_\varphi$ and
   $g_d$ as free parameters and examine responses to their
   variation. Here we fix $M^2_{\varphi} = 0.1 \Lambda^2$ and change
   $g_d$.  The phase structure is determined by the relative strength
   $g_v/M^2_{\varphi}$. For example, the ordinary results of the QM
   model are reproduced by taking $g_v/M^2_{\varphi} \rightarrow 0$.
   Our choice of the parameter set is summarized in
   Table~\ref{tab:parameter_frg}.

   \subsection{Phase structure}
   \label{sec:phase-structure} Before studying the critical region, we
   start with an analysis of the phase structure with varying $g_d$.
   \begin{figure}[!ht]
    \centering \includegraphics[width=0.49\columnwidth]{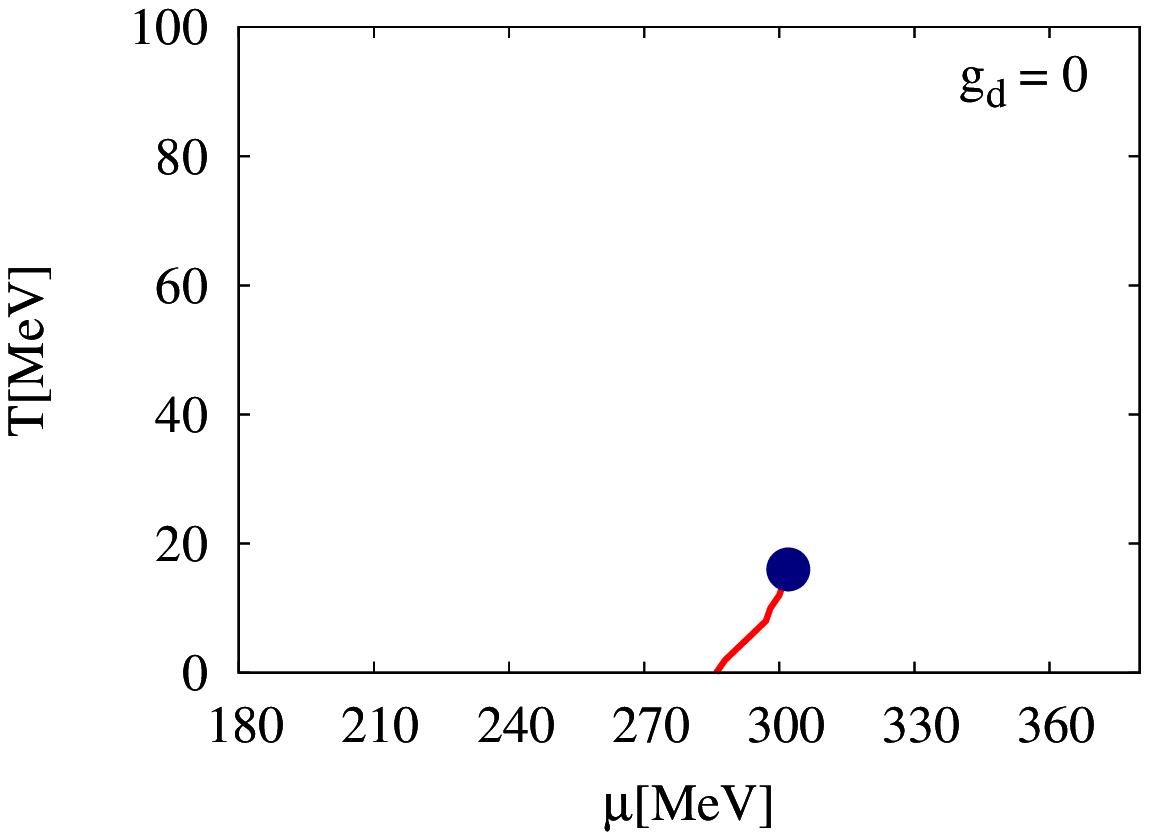}
    \includegraphics[width=0.49\columnwidth]{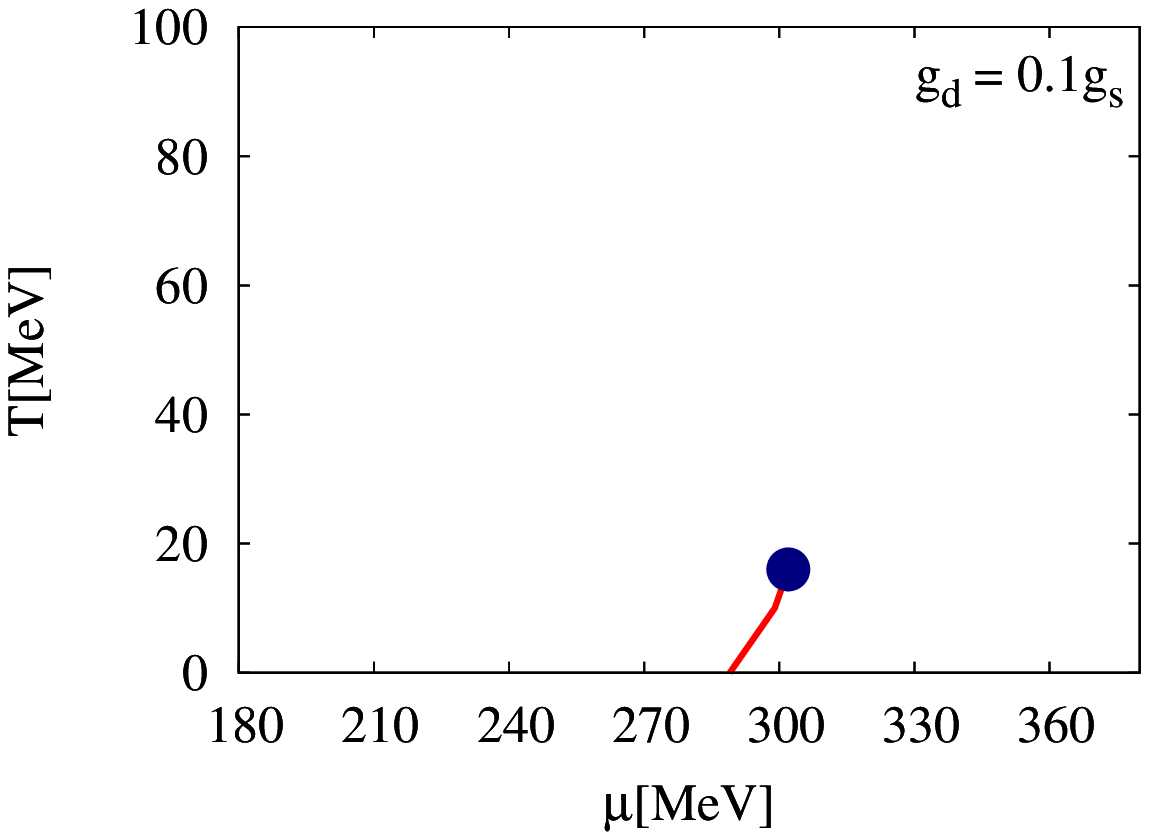}
    \includegraphics[width=0.49\columnwidth]{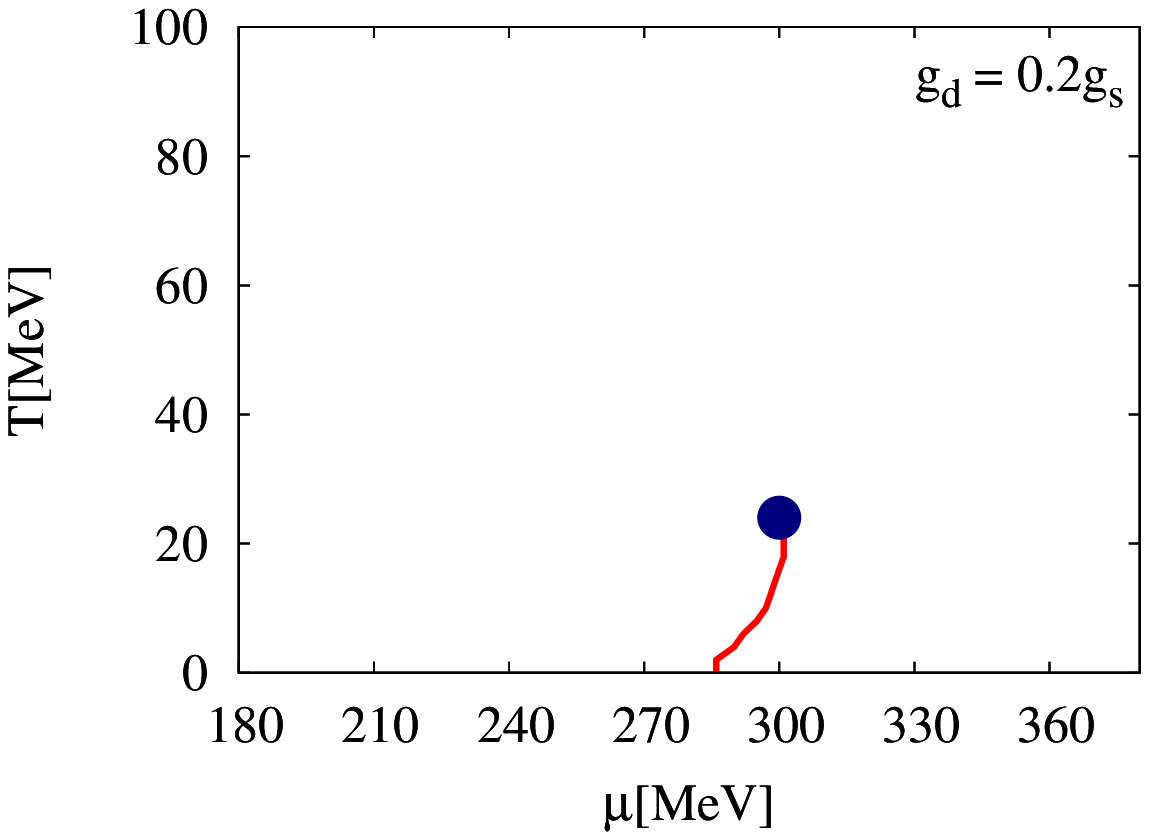}
    \includegraphics[width=0.49\columnwidth]{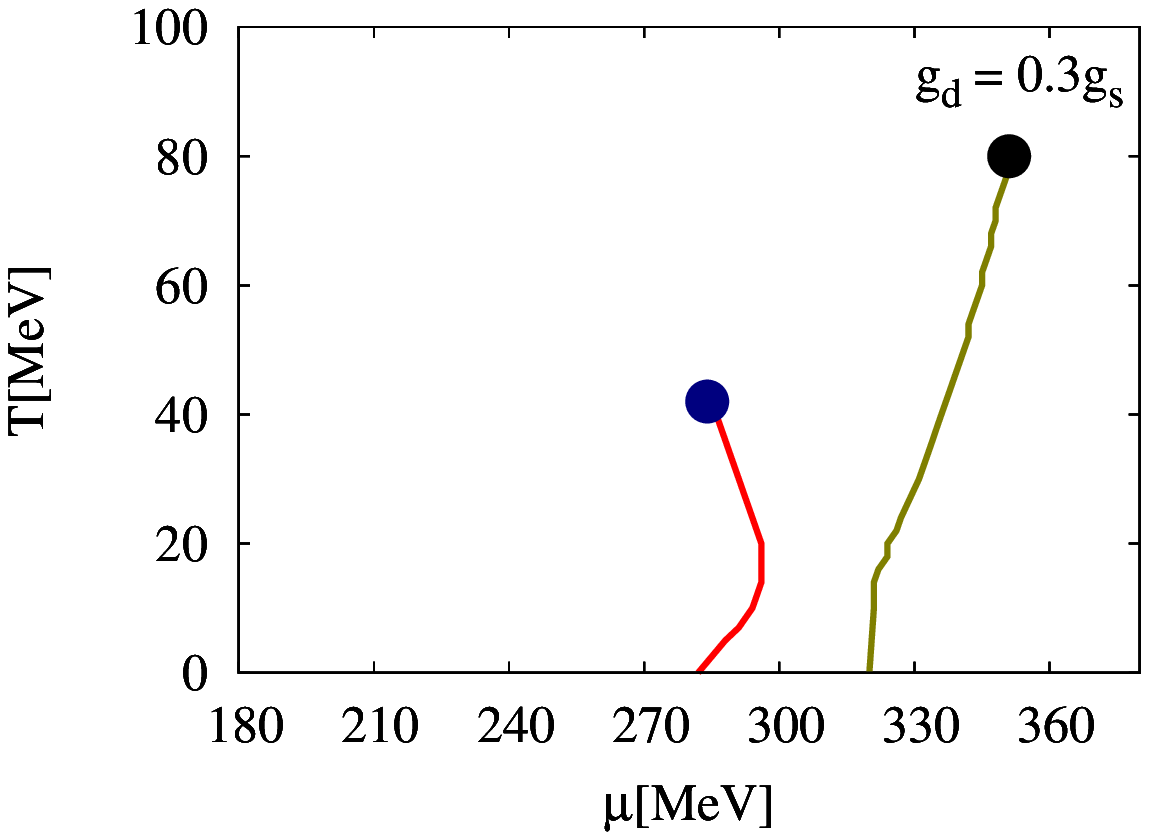}
    \caption{Phase diagram for four different $g_d$ ($0$, $0.1g_s$,
    $0.2g_s$ and $0.3g_s$).  The solid line denotes the first-order
    phase
    boundary while the blue closed circle represents the critical
    point. A green line and black circle in $g_d = 0.3g_s$ denote
    second first-order phase boundary and its critical end point.}
    \label{fig:phase_diagram}
   \end{figure}
   In Fig.~\ref{fig:phase_diagram}, the phase diagrams are shown for
   four fixed values of $g_d$.  In the present work, we have generalized
   the previous results in the zero density-coupling case
   \cite{Kamikado:2012bt} to the finite couplings.  We find the QCD
   critical point for each $g_d$ value. The locations of the QCD
   critical point are, in unit of MeV, at ($T_{cp}$,$\mu_{cp}$) $\simeq$
   $(15,303),(16,303),(28,300)$ and $(42,284)$ for $g_d =
   0,0.1g_s,0.2g_s$ and $0.3 g_s$, respectively.  For the temperature
   below the critical point, we find the first-order phase boundary
   which separates a chiral restored phase (higher $\mu$) and a broken
   phase (lower $\mu$). At the critical point, the phase transition is
   second order.  Beyond the critical point, the phase transition
   becomes crossover and the pseudo critical line is not shown in the
   phase diagram.

   The position of the critical point slightly rises to a higher
   temperature direction with increasing $g_d$. This is an opposite
   behavior to the well known results in the NJL model with the vector
   interaction
   \cite{Kitazawa:2002bc,Sasaki:2006ws,Sakai:2008ga,Fukushima:2008is}.
   This is because our effective model corresponds to the attractive
   vector interaction case ($G_V < 0$). In fact, a similar behavior is
   found in the mean field approximation of the NJL model with an
   artificial attractive vector interaction \cite{Fukushima:2008wg}.

   In the figure for $g_d = 0.3 g_s$, one sees another critical point at
   ($T_{cp}$,$\mu_{cp}$) $\simeq$ (80,351) MeV from which another
   first-order phase boundary is extended to the lower-$T$ direction.
   This should be an artifact of our model. This additional critical
   point and its associated first-order phase boundary appear only with
   large couplings such as $g_d =0.3 g_s$ and they belong to a different
   branch from the QCD critical point.  The magnitude of the chiral
   condensate is sufficiently smaller than the vacuum expectation value
   and jumps from about 10 MeV to a few MeV across the first-order phase
   boundary.  At $g_d = 0.3 g_s$, the attractive force between two
   quarks is too strong and causes another stable minimum of the
   effective potential in the high density and small chiral condensation
   region.
   \begin{figure}[!ht]
    \centering \includegraphics[width=0.49\columnwidth]{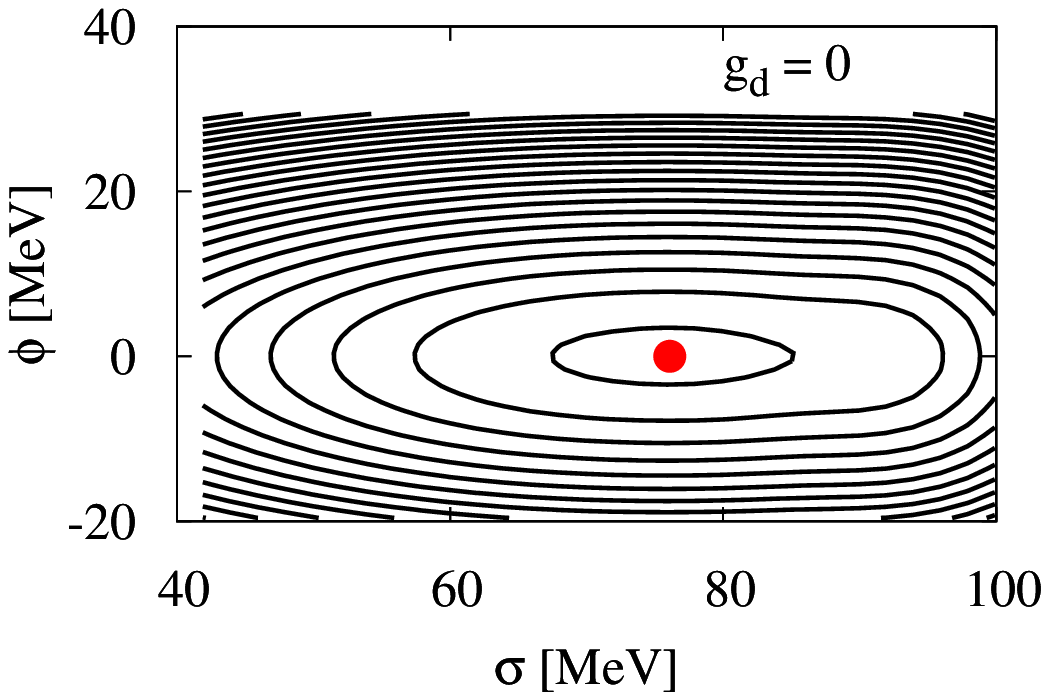}
    \includegraphics[width=0.49\columnwidth]{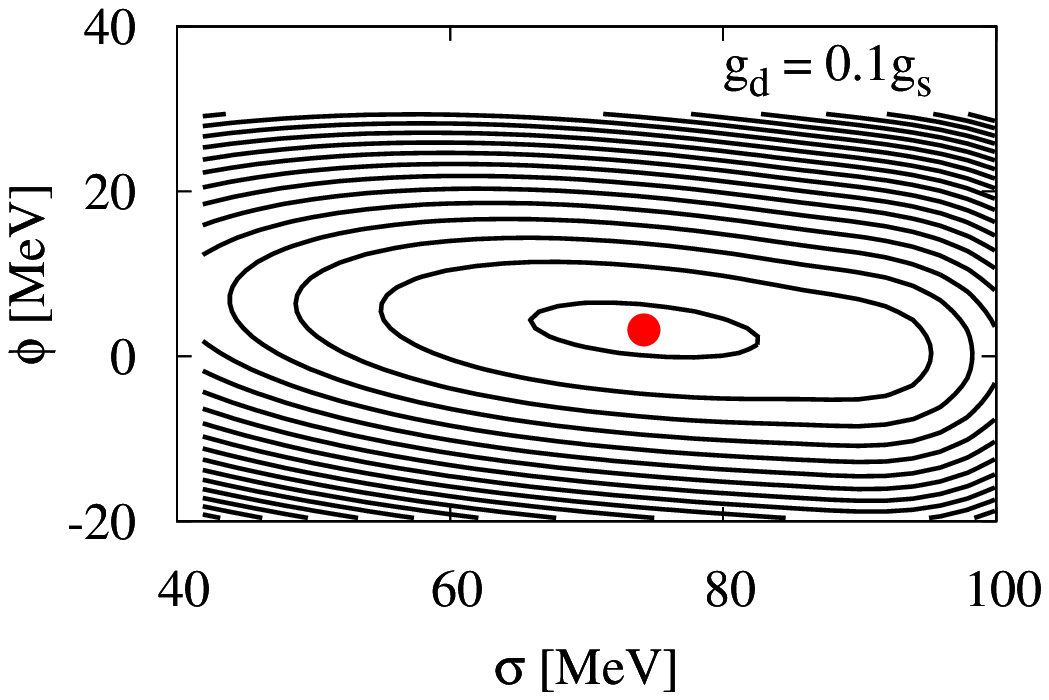}
    \includegraphics[width=0.49\columnwidth]{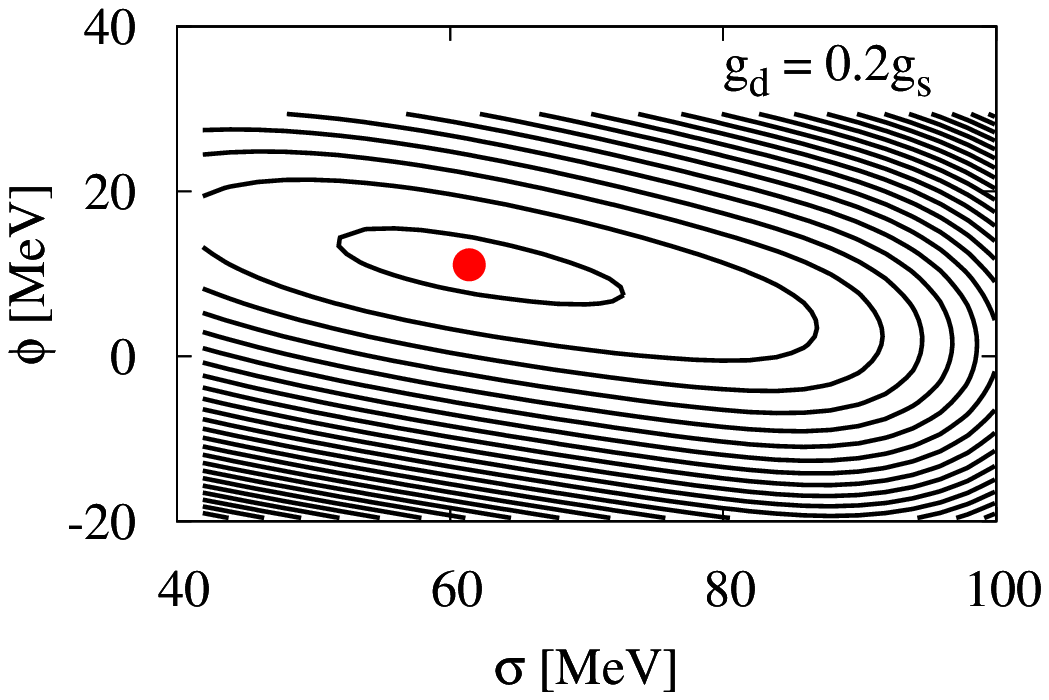}
    \includegraphics[width=0.49\columnwidth]{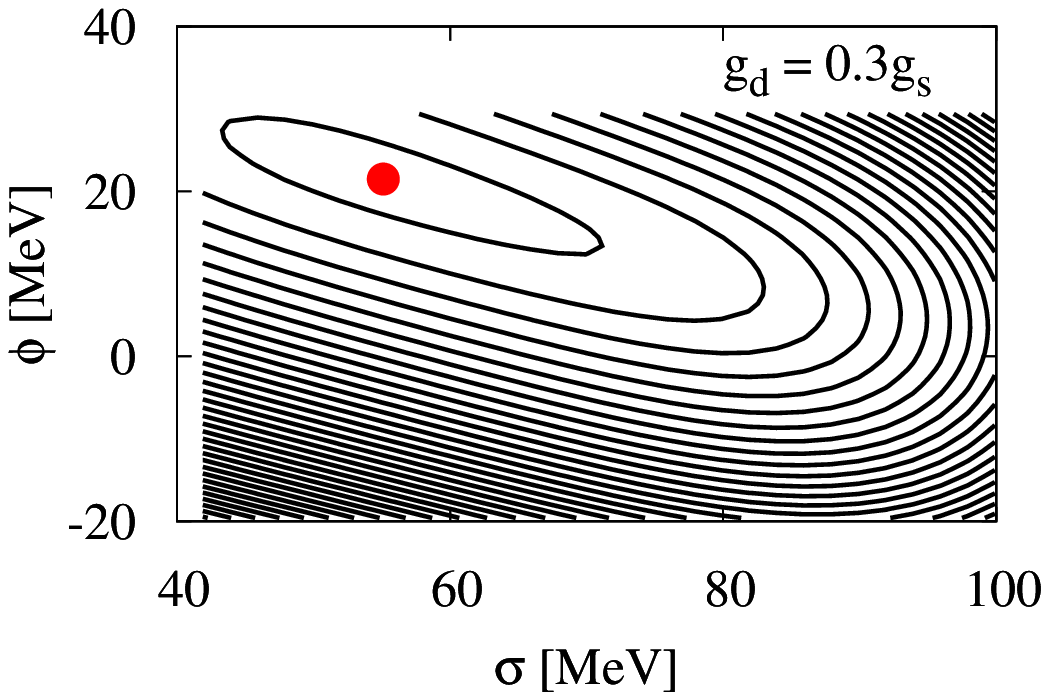}
    \caption{Contours of the effective potential in the $\sigma$ and
    $\varphi$ space near the critical point for varying $g_d$. The solid
    points show the minimum of the effective potential.}
    \label{fig:potential}
   \end{figure}
   In Fig.~\ref{fig:potential}, the effective potential contour is shown
   for several $g_d$ values at temperature and chemical potential close
   to the critical point. Since the phase transition is second order at
   the QCD critical point, there is a flat direction on the effective
   potential around the minimum.

   For $g_d = 0$, $\varphi$ is decoupled from the chiral modes then the
   mixing term such as $\sigma^2 \varphi$ does not arise in the
   effective potential.  As a result, we can see the flat direction is
   parallel to the $\sigma$ axis.

   On the other hand, for $g_d \neq 0$, there is a mixing between
   $\sigma$ and $\varphi$ and the expectation value of $\varphi$ is not
   zero.

   The flat direction is tilted to the direction of the linear
   combination of $\sigma$ and $\varphi$ directions and the $\sigma$
   direction is no longer special.  These behaviors are also found in a
   calculation on the NJL model \cite{Fujii:2004jt}.  The angle between
   the flat direction and the $\sigma$ direction becomes larger with
   increasing $g_d$ as the mixing between the $\sigma$ and $\varphi$
   becomes stronger.

   The singular behavior of susceptibilities shown in the next
   section is mediated by the fluctuation along the flat direction of
   the effective potential. For finite $g_d$, the flat direction is the
   linear combination of the $\sigma$ and $\varphi$ directions and the
   critical behaviors are attributed to the fluctuation of the linear
   combination. On the other hand, in the case of $g_d = 0$ (the QM
   model), the flat direction is parallel to the $\sigma$ axis and the
   critical behaviors are attribute to only the fluctuation of the
   chiral order parameter.

   \subsection{Susceptibility}
   \label{sec:susceptibility} The information on the size of the
   critical region is important for experiments in search of the QCD
   critical point.  We define the critical region by using the
   quark-number susceptibility $\chi_q$ given as the response of the
   quark-number density $n_q$ to $\mu$:

   \begin{equation} \chi_q \equiv
    \frac{\partial n_q}{\partial \mu} = \frac{\partial^2
     P(T,\mu)}{\partial \mu^2} \;,p
   \end{equation}
   where the pressure $P$ is given by $P(T,\mu) = -U(\sigma_0,\phi_0)$.
   For the comparison of the shape and size of the critical region for
   different $g_d$, it is convenient to use a normalized susceptibility
   $\chi_q^{\rm{n}}$, i.e., the ratio to that of the mass-less free
   quark gas $\chi_q^{\rm free}$:
   \begin{equation}
    \begin{split}
     \chi_q^{\rm free} & \equiv \frac{2 N_c N_f }{6} \left[T^2 +
     \frac{3 \mu^2}{ \pi^2}\right] \\
     \chi_q^{\rm{n}} &\equiv \chi_q / \chi_q^{\rm free}   \;.\\
    \end{split}
   \end{equation}
   \begin{figure}[!h]
    \centering \includegraphics[width=0.49\columnwidth]{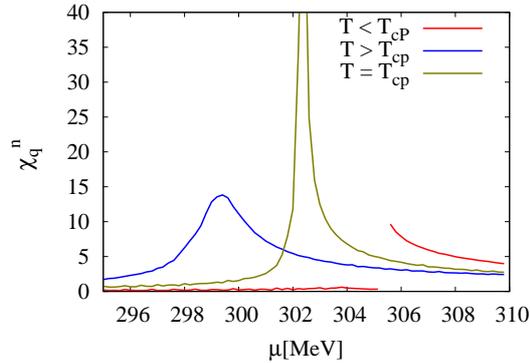}
    \caption{Plots of the normalized quark-number susceptibility
    $\chi_q^{\rm{n}} = \chi_q / \chi_q^{\rm free}$ with chemical
    potential at $g_d = 0.2g_s$. The red/blue line represent the
    $\chi_q^{\rm{n}}$ at T below/beyond the $T_{cp}$. The green line is
    at $T = T_{cp}$.}  \label{fig:2d_sus}
   \end{figure}
   
   In Fig.~\ref{fig:2d_sus}, $\chi_q^{\rm{n}}$ at $g_d = 0.2 g_s$ is
   shown as a function of quark-chemical potential for three different
   temperatures.  For temperatures below the critical temperature
   $T_{cp}$, there is a discontinuity in $\chi_q^{\rm{n}}$ at the
   first-order phase boundary.  For temperatures above $T_{cp}$,
   $\chi_q^{\rm{n}}$ is continuous as a function of $\mu$ and has a peak
   at the crossover phase boundary. The peak height becomes higher with
   $T$ approaching to $T_{cp}$.  At $T = T_{cp}$, the peak height of
   $\chi_q^{\rm{n}}$ diverges.  Near the QCD critical point, we find the
   same behavior of the $\chi_q^{\rm{n}}$ for all $g_d$.
   
   \begin{figure}[!ht] \centering
    \includegraphics[width=0.49\columnwidth]{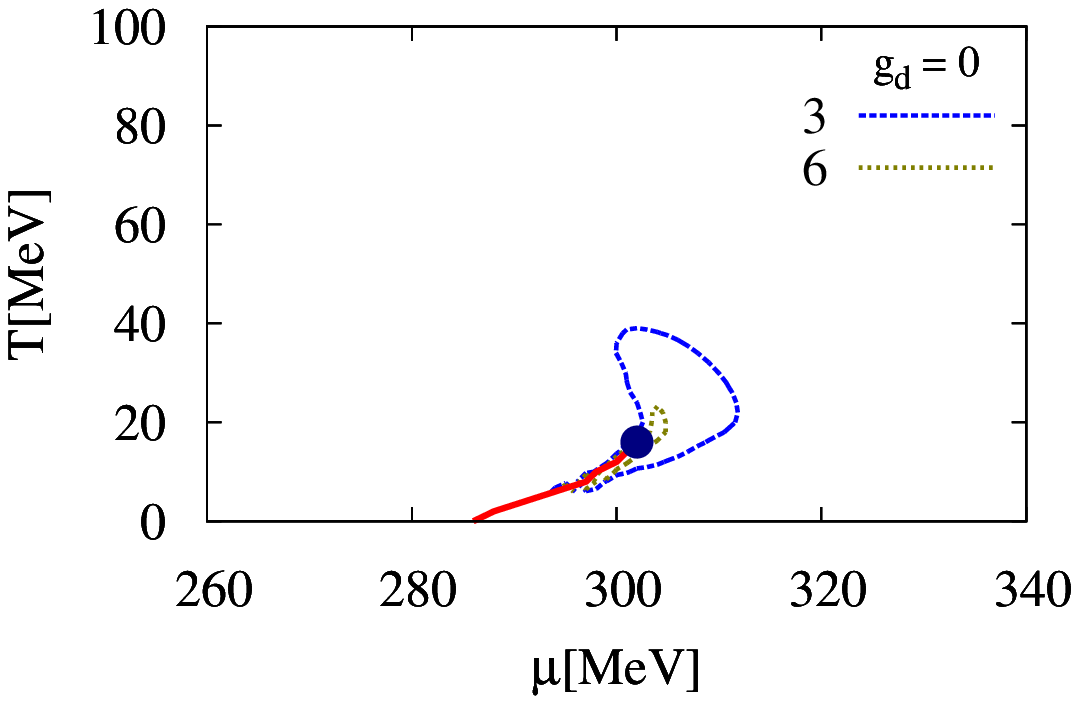}
    \includegraphics[width=0.49\columnwidth]{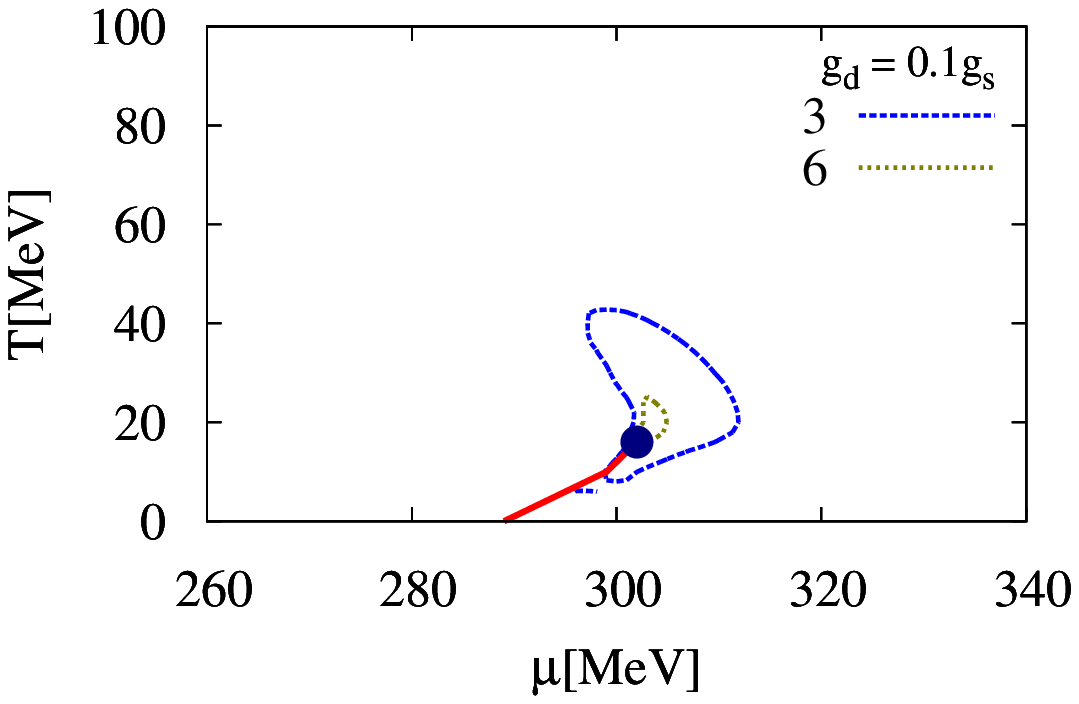}
    \includegraphics[width=0.49\columnwidth]{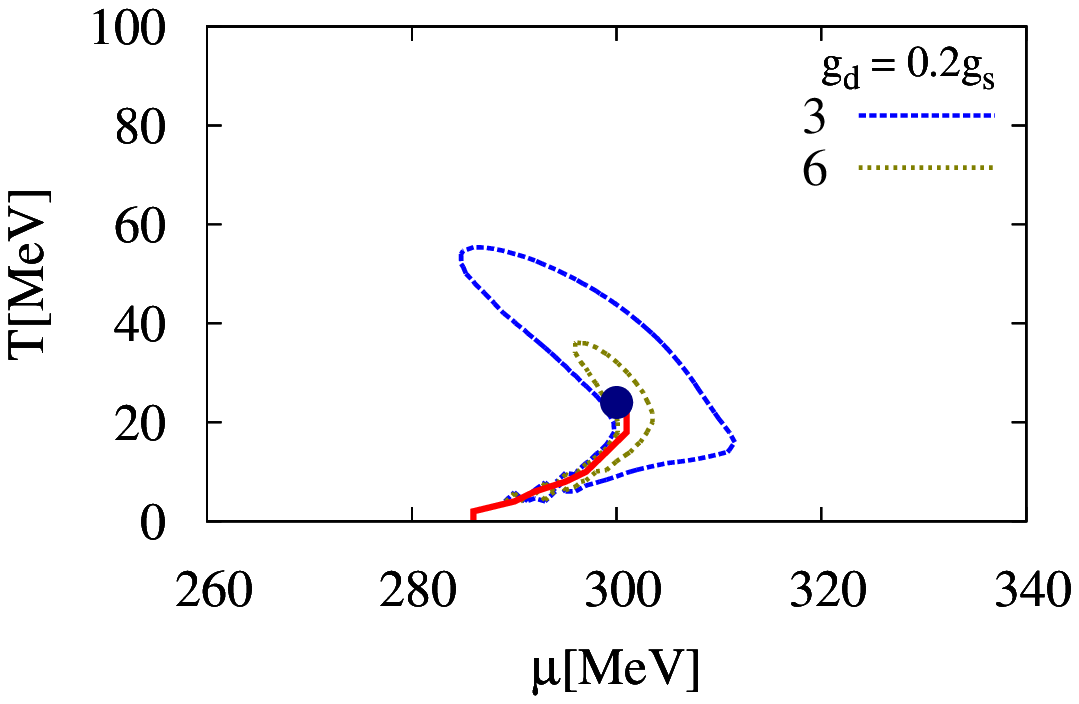}
    \includegraphics[width=0.49\columnwidth]{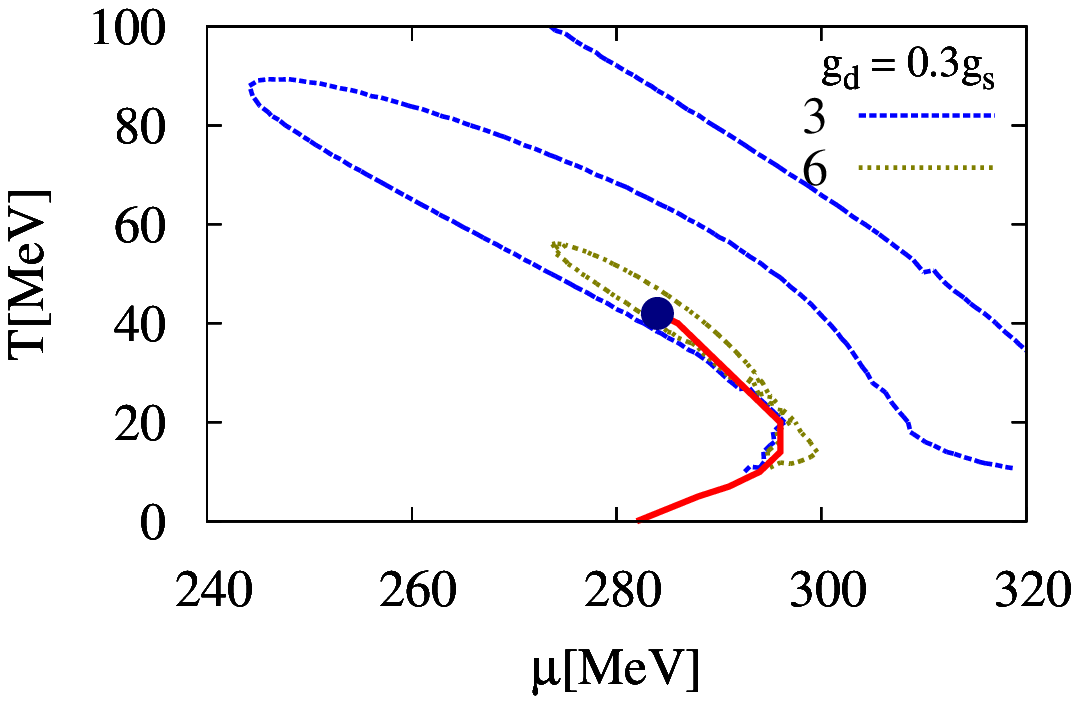}
    \caption{Contours of the normalized quark-number susceptibility
    $\chi_q^{\rm{n}} = \chi_q / \chi_q^{\rm free}$ around the critical
    point for $\chi_q^{\rm{n}} = 3$ and $6$.  The red line denote the
    first-order phase boundary. The blue solid circle represent the QCD
    critical point. } \label{fig:susceptibility}
   \end{figure}
   In Fig.~\ref{fig:susceptibility}, the contour plots of
   $\chi_q^{\rm{n}}$ with two fixed values, $\chi_q^{\rm{n}}=3,6$, are
   shown for varying $g_d$. The criticality near the QCD critical point
   is characterized by the enhancement of $\chi_q$.  In the following,
   we define the critical region by $\chi_q^{\rm{n}} > 3$.

   For $g_d = 0$, we have the critical region which has 10 MeV extent
   both in the $T$ and $\mu$ directions.  The size of the critical
   region for $g_d = 0$ is consistent with the previous work
   \cite{Schaefer:2006ds} and the width of the critical region is
   shrunken in the $\mu$ direction by almost one order of magnitude
   compared with the mean-field calculation \cite{Schaefer:2006ds}.

   For finite $g_d$, the critical region is deformed from that at $g_d =
   0$.  We observe the expansion of the critical region to the crossover
   phase transition direction with increasing $g_d$.  The shape and size
   of the critical region are quite sensitive to $g_d$ such that it
   shows drastic deformation while the position of the QCD critical
   point moves only slightly.  At $g_d = 0.3 g_s$, the critical region
   is elongated to the cross over side twice as large as that for $g_d =
   0$.

   This expansion can be partly understood by looking at the curvature
   masses of the linear combination of the $\sigma$ and $\varphi$, as
   demonstrated in the following.
	  
  \subsection{curvature masses}
  \label{sec:curvature-masses} 
\begin{figure}[!h] \centering
   \includegraphics[width=0.49\columnwidth]{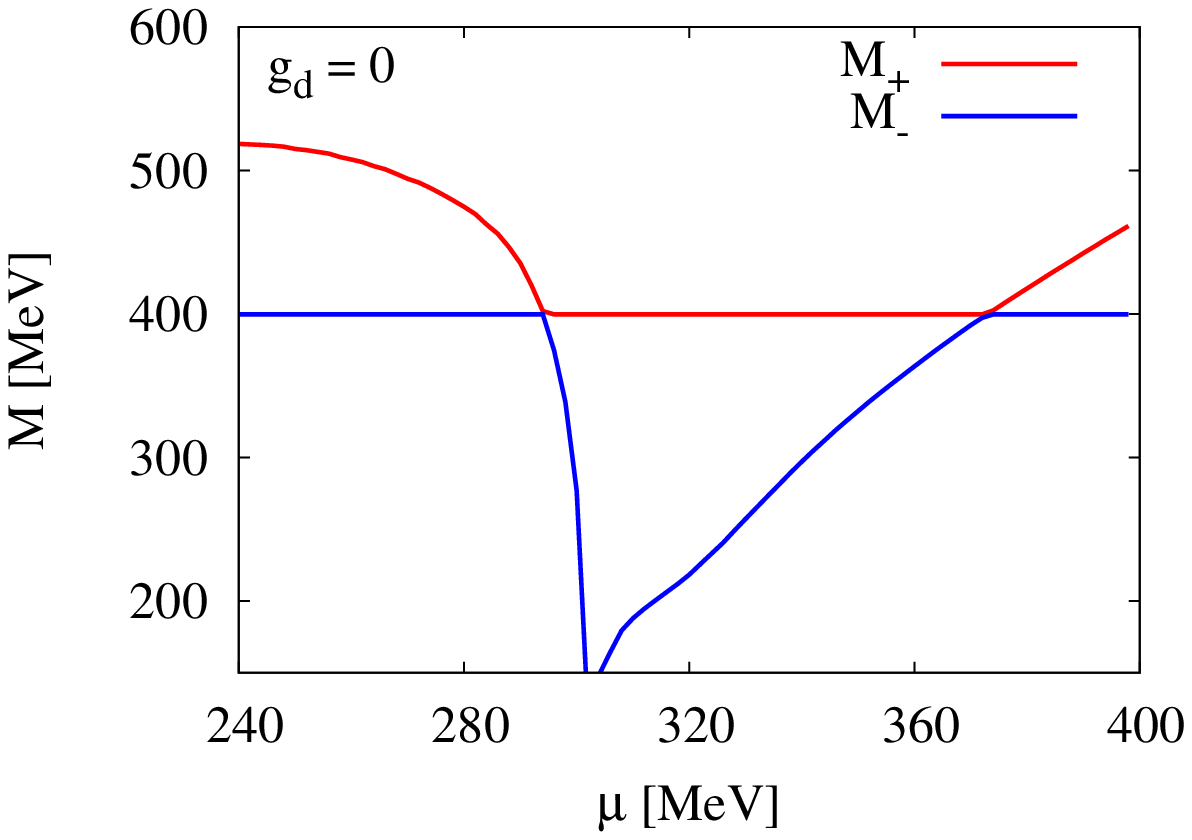}
   \includegraphics[width=0.49\columnwidth]{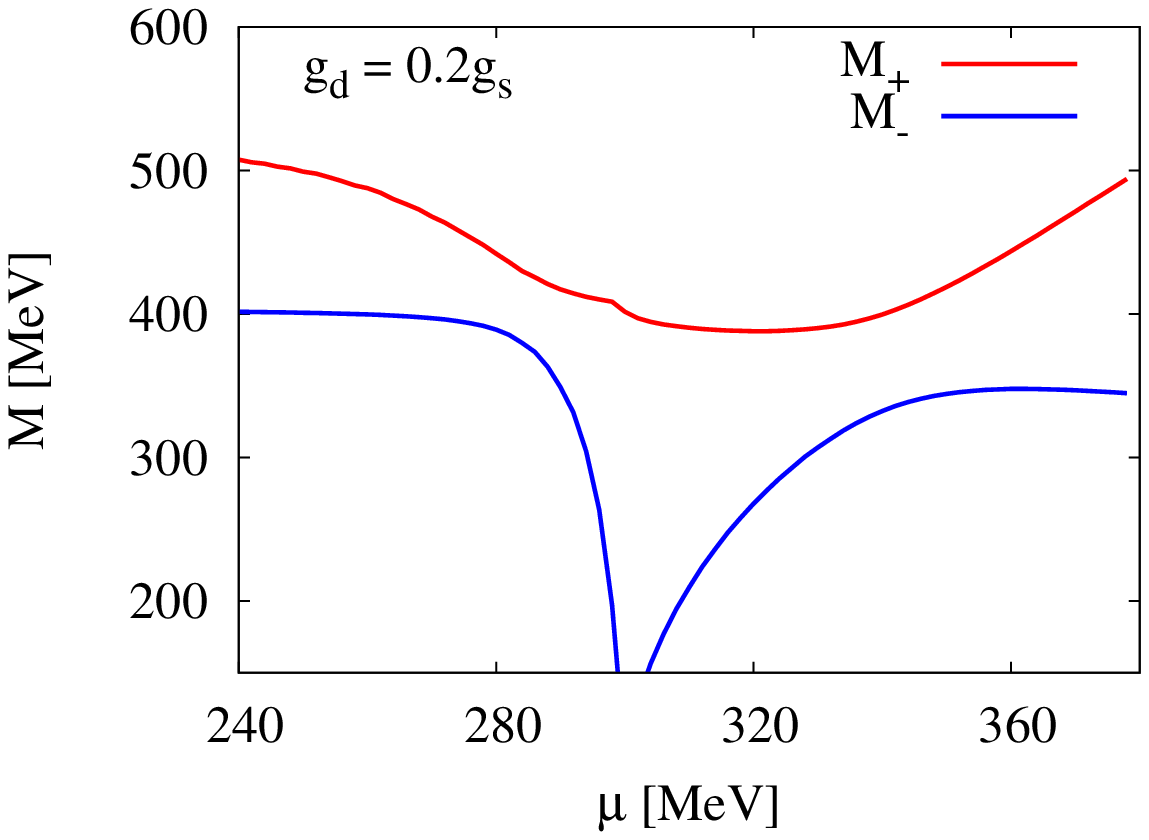}
   \caption{Plots of the curvature masses $M_{\pm}$ near the critical
   point with chemical potential. The left and right panel are at $g_d =
   0$ and $g_d = 0.2g_s$ respectively.}  \label{fig:2d_mass}
  \end{figure}

In order to give the understanding of the
  expanding behavior of the critical region, we calculate the curvature
  masses of the linear combination of the $\sigma$ and $\varphi$.  The
  masses $M_{+}$ and $M_{-}$ are defined as a bigger and smaller
  eigenvalue of the curvature matrix, respectively:
  \begin{equation}
   M = 
    \begin{pmatrix}
     \frac{\partial^2 U}{\partial \sigma \partial \sigma}&
     \frac{\partial^2 U}{\partial \sigma \partial \varphi} \\ 
     \frac{\partial^2 U}{\partial \varphi \partial \sigma} &
     \frac{\partial^2 U}{\partial \varphi \partial \varphi} 
    \end{pmatrix} 
    \Bigg|_{ \sigma_0,\varphi_0}
    \;.
    \label{eq:curvature_matrix}
  \end{equation}
  In Fig.~\ref{fig:2d_mass}, the curvature masses $M_{\pm}$ at
  $T=T_{cp}$ are shown as functions of quark-chemical potential for $g_d
  = 0$ (left) and $g_d = 0.2g_s$ (right).
  
  At $g_d = 0$, there is no mixing between the $\sigma$ and $\varphi$
  then the non-diagonal parts of the mass matrix
  (\ref{eq:curvature_matrix}) are always zero, and thus we can identify
  the $\sigma$ and $\varphi$ modes individually.  Since the $\varphi$ is
  decoupled from the matter sector, the curvature mass of the $\varphi$
  is constant ($\sim 400 $ MeV) with chemical potential varied while the
  sigma mass decreases below the mass of the $\varphi$ and vanishes at
  the critical point.  Beyond the critical point, the $\sigma$ mass
  increases monotonically.

  In contrast, the non-diagonal parts of the mass matrix
  (\ref{eq:curvature_matrix}) do not vanish for $g_d \neq 0$ so that the
  $\sigma$ and $\varphi$ are mixed.  At $g_d = 0.2 g_s$, $M_-$ stays
  constant ($\sim 400$ MeV) while $M_+$ decreases below the critical
  chemical potential.  When $M_+$ gets close to $M_-$, $M_-$ starts to
  decrease and vanishes at the critical point. We observe a kind of
  level repulsion near the critical point between $M_+$ and $M_-$ which
  is familiar in a perturbation theory in quantum mechanics
  \cite{Schiff:1968ev}.  Beyond the critical point, $M_-$ increases
  until 340 MeV and becomes almost constant. $M_+$ increases again while
  $M_-$ increases and approaches a constant value. We found these
  behaviors of the $M_+$ and $M_-$ generally for $g_d \neq 0$.

  Note that the softening behavior shown here does not correspond to the
  dynamical softening of these modes.  In order to see the dynamical
  softening of the modes, one has to calculate spectral functions of the
  $\sigma$ or $\varphi$ modes in which one would find a dissipative zero
  mode in the space-like region of the spectral function
  \cite{Fujii:2003bz,Fujii:2004jt,Son:2004iv,Minami:2009hn}.
  
  \begin{figure}[!h] \centering
   \includegraphics[width=0.49\columnwidth]{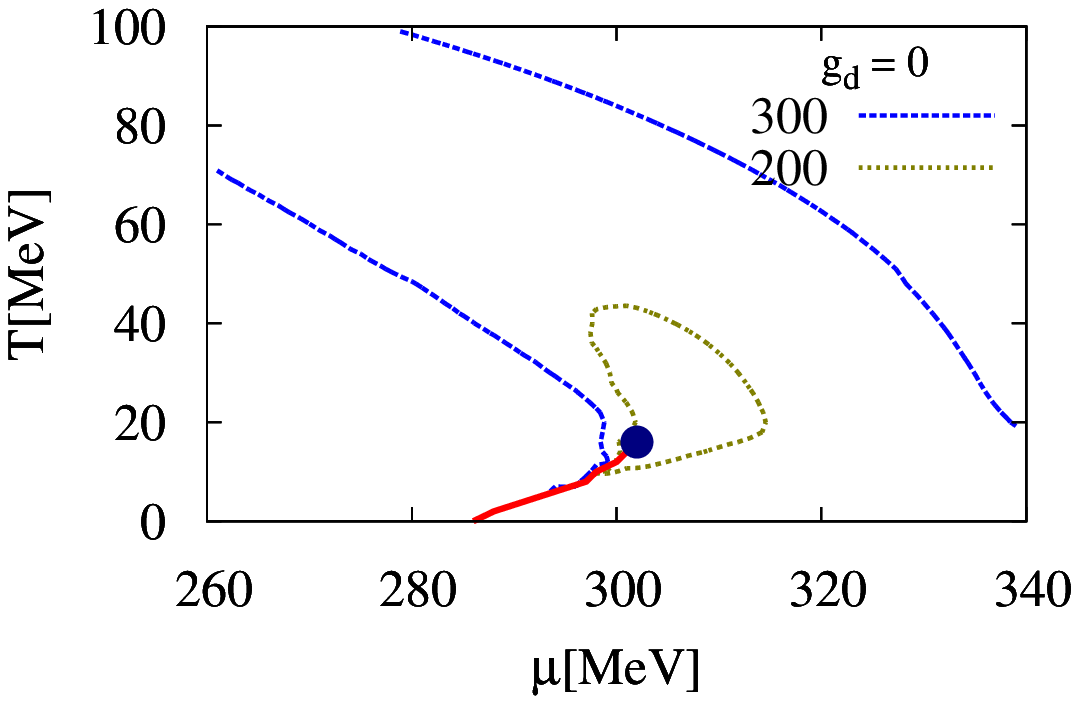}
   \includegraphics[width=0.49\columnwidth]{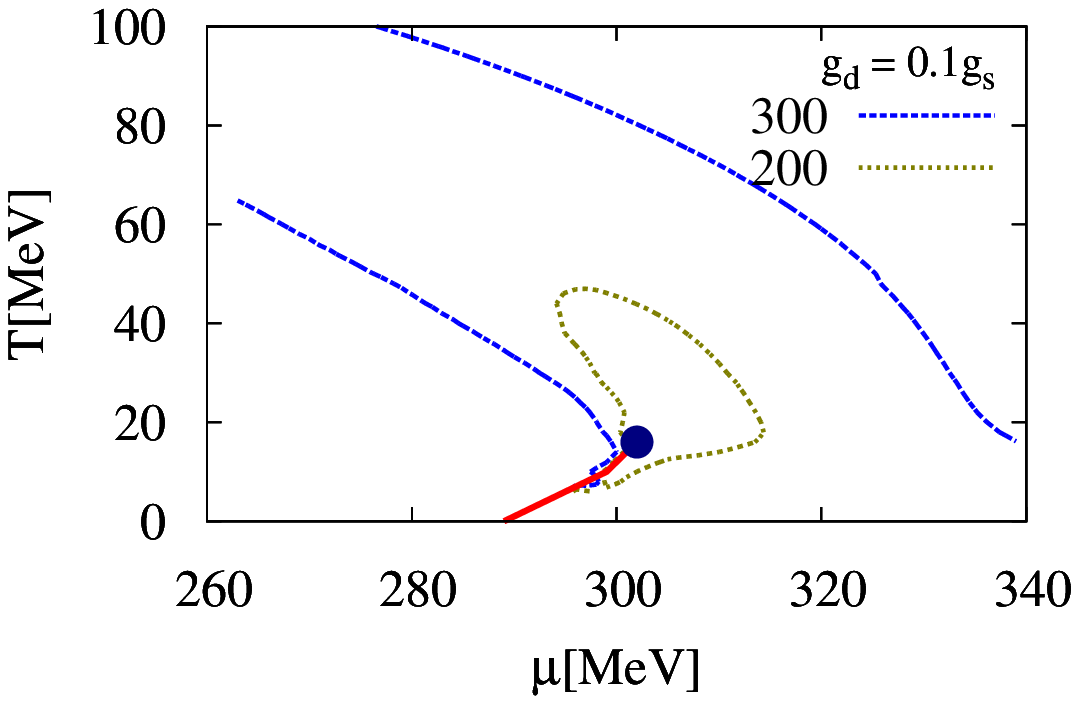}
   \includegraphics[width=0.49\columnwidth]{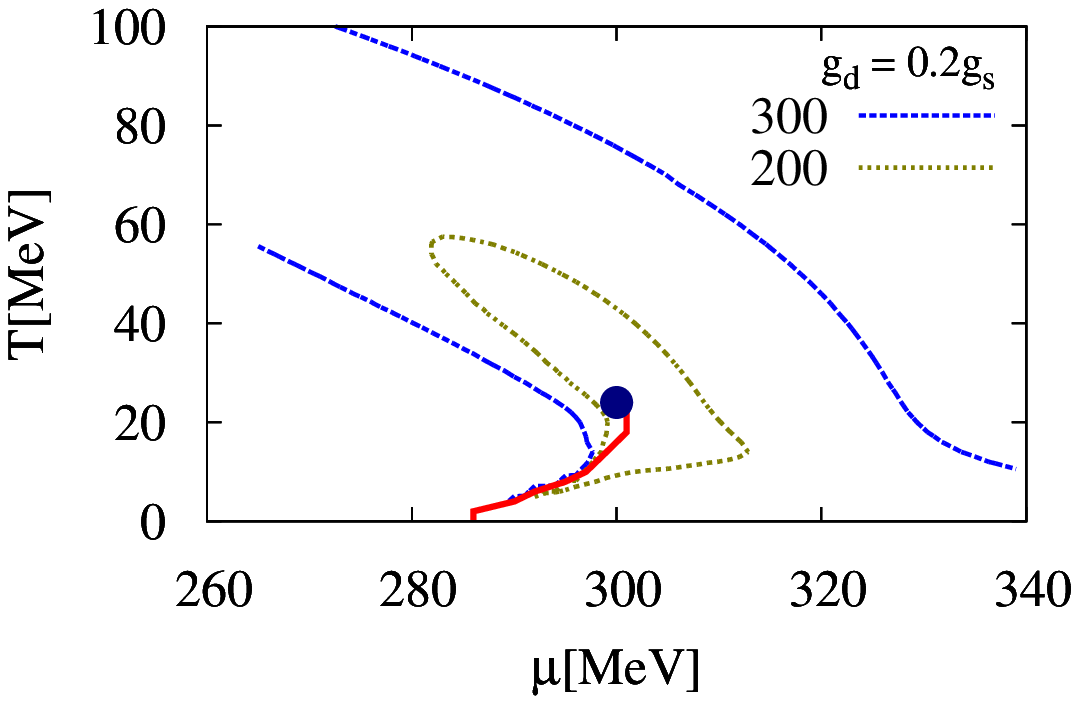}
   \includegraphics[width=0.49\columnwidth]{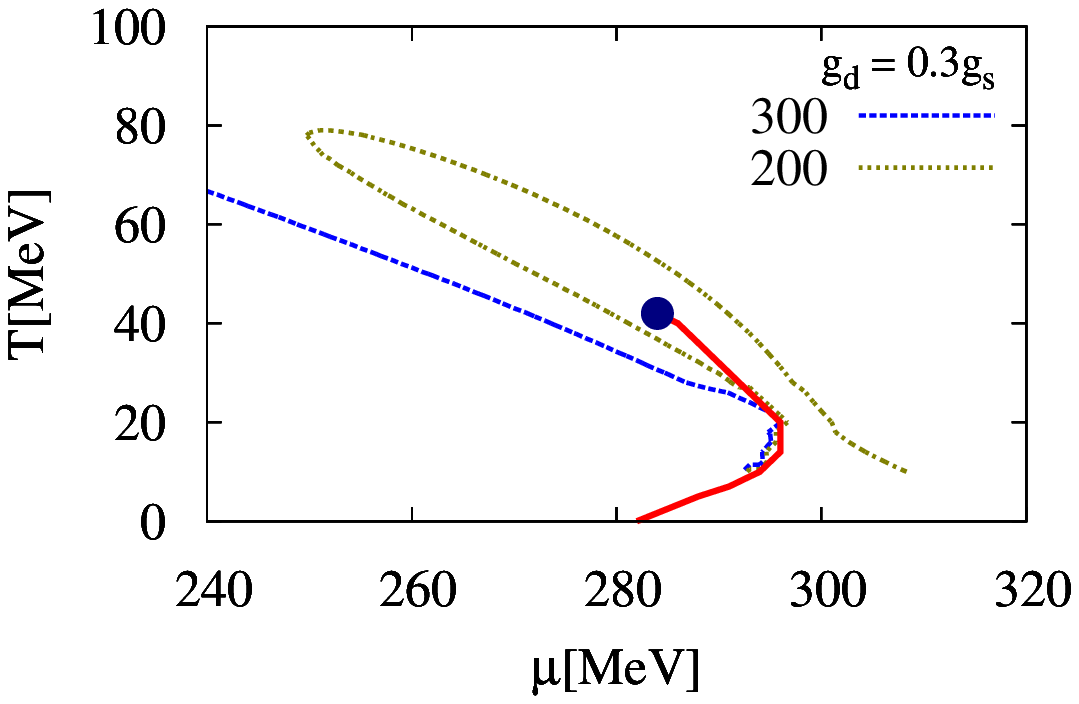}
   \caption{Contours of the smaller curvature mass $M_-$ for varying
   $g_d$. The contours are shown for $M_- = 200$ and $300$ MeV. The red
   lines denote the first-order phase boundary. The blue solid circles
   represent the QCD critical point. } \label{fig:curvature}
  \end{figure}
  In Fig~\ref{fig:curvature}, we show a contour map for $M_-$ with
  varying $g_d$.  For each $g_d$, $M_-$ is close to zero in the vicinity
  of the critical point. The softening of the linear combination of the
  $\sigma$ and $\varphi$ causes the critical behavior on the
  susceptibility.

  In the following, we call the area in which $M_- < 200$ MeV the
  softening region. For each $g_d$, we can see the softening region
  almost coincide with the critical region, i.e., the region in which
  $\chi_q^{\rm{n}} > 3$ in Fig.~\ref{fig:susceptibility}.  The softening
  region is also expanded to the crossover phase transition direction
  with $g_d$ being increased.

  The level repulsion of the curvature masses precisely corresponds to
  the expanding behavior of the critical region.  The curvature mass
  $M_-$ is reduced when the density coupling or the non-diagonal parts
  of the mass matrix~(\ref{eq:curvature_matrix}) is finite.  The
  non-diagonal parts of (\ref{eq:curvature_matrix}) become larger as
  $g_d$ increases. As a result, the level repulsion between $M_+$ and
  $M_-$ becomes larger and the region with small $M_-$ also gets
  expanded.  It simply follows that the inclusion of the mixing between
  the $\sigma$ and $\varphi$ and does not depend on the details of
  interactions, for example, repulsive or attractive force.

  \section{Summary} \label{sec:summary} We have examined the influence
  of the mixing between the chiral condensation and the quark-number
  density on the critical region around the QCD critical point.  The
  fluctuations of the quark-number density as well as the chiral modes
  were taken into account through the functional renormalization group
  (FRG) equation.

  We have extended the quark meson model to a new effective model which
  is appropriate for the correct description of the dynamics near the
  QCD critical point. In addition to the quarks and chiral modes
  ($\sigma$ and $\vec{\pi}$), the new effective model contains a new
  field $\varphi$ which corresponds to the quark-number density. The
  magnitude of the mixing between the $\sigma$ and $\varphi$ was
  determined by a density couping $g_d$.

  We derived the flow equation for the scale dependent effective action
  with the fluctuation of the density field at a finite temperature and
  quark-chemical potential.  The flow equation for the scale dependent
  effective potential was solved numerically on a two-dimensional grid
  of the chiral condensate ($\sigma$) and quark-number density
  ($\varphi$).

  We found that the QCD critical point exists for varying $g_d$ while
  its position slightly moves to the higher temperature and lower
  quark-chemical potential direction as $g_d$ increases. The
  quark-number susceptibility was enhanced around the critical point.

  In contrast, we have shown the shape and size of the critical region
  was quite sensitive to the magnitude of the mixing term.  We have
  shown that the critical region is enlarged to a lower chemical
  potential and higher temperature direction when we increase the
  density coupling $g_d$.

  In order to confirm the expanding behavior, we also calculated the
  curvature masses of the linear combination of the $\sigma$ and
  $\varphi$. It turned out that the critical region was almost identical
  to the softening region where the curvature of the effective potential
  becomes small. The softening of the linear combination of the $\sigma$
  and $\varphi$ causes the critical behavior.

  We have shown that the origin the expanding behavior of the critical
  region and the softening region 
  level repulsion of the curvature masses given by the linear
  combination of the chiral condensate and the quark-number density. The
  large level repulsion at large $g_d$ makes the critical and softening
  region wider. The level repulsion is a generic consequence of the
  mixing between the chiral condensate and the quark-number density.

  As a future work, it is interesting to investigate higher moments of
  the quark-number density or the charge-number density including the
  quark-density coupling and the fluctuation of the quark-number
  density. The higher moments are considered to show the stronger
  criticality \cite{Stephanov:2008qz} and would be more sensitive to the
  mixing than the quark-number susceptibility.  In the present work, we
  have considered the simplest case of the coupling and treated its
  strength as a free parameter but it should be determined by a more
  microscopic model or experiments. The exact determinations of the
  strength or detailed form of the mixing also awaits future studies.

 \section*{Acknowledgements}
 \label{sec:acknowledgements} K.~K. was supported by Grant-in-Aid for
 JSPS Fellows (No.22-3671).  The numerical calculations were carried out
 on SR16000 at YITP in Kyoto University.  This work is supported in part
 by the Grants-in-Aid for Scientific Research from JSPS and and MEXT
 (Nos. 
	  23340067, 
	  K. Morita, T. Kunihiro) and Innovative Areas (No. 2404:
	  24105001, 24105008) 
	  International Program for Quark-hadron Sciences, and by the
	  Grant-in-Aid for the global COE program ``The Next Generation
	  of Physics, Spun from Universality and Emergence" from MEXT.
	  

 \end{document}